\documentclass[compsoc, conference, a4paper, 10pt, times]{IEEEtran}

\usepackage{soul} 
\usepackage{array}
\usepackage{comment}
\usepackage{hyperref}
\usepackage{graphicx}

\sethlcolor{lime}
\newcommand{\todo}[2][]{
    \textcolor{red}{\hl{\textbf{#1} #2}}
}

\graphicspath{{./}{figures/}}

\begin{document}

\title{Enhancing Vulnerability Prioritization: Data-Driven Exploit Predictions with Community-Driven Insights}


\author{
\IEEEauthorblockN{Jay Jacobs}
\IEEEauthorblockA{\textit{Cyentia Institute} \\
jay@cyentia.com}
\and
\IEEEauthorblockN{Sasha Romanosky}
\IEEEauthorblockA{\textit{RAND Corporation} \\
sromanos@rand.org}
\and 
\IEEEauthorblockN{Octavian Suciu}
\IEEEauthorblockA{\textit{University of Maryland} \\
osuciu@umd.edu}
\and
\IEEEauthorblockN{Ben Edwards}
\IEEEauthorblockA{\textit{Cyentia Institute} \\
ben@cyentia.com}
\and
\IEEEauthorblockN{Armin Sarabi}
\IEEEauthorblockA{\textit{University of Michigan} \\
arsarabi@umich.edu}
}

\ifodd 1
    \newcommand{\com}[1]{\textbf{\color{red}(COMMENT: #1)}}
    \newcommand{\rev}[1]{{\color{blue}#1}}
\else
    \newcommand{\com}[1]{}
    \newcommand{\rev}[1]{#1}
\fi

\maketitle

\begin{abstract}
The number of disclosed vulnerabilities has been steadily increasing over the years. At the same time, organizations face significant challenges patching their systems, leading to a need to prioritize vulnerability remediation in order to reduce the risk of attacks.

Unfortunately, existing vulnerability scoring systems are either vendor-specific,  proprietary, or are only commercially available. Moreover, these and other prioritization strategies based on vulnerability severity are poor predictors of actual vulnerability exploitation because they do not incorporate new information that might impact the likelihood of exploitation.

In this paper we present the efforts behind building a Special Interest Group (SIG) that seeks to develop a completely data-driven exploit scoring system that produces scores for all known vulnerabilities, that is freely available, and which adapts to new information. The Exploit Prediction Scoring System (EPSS) SIG consists of more than 170 experts from around the world and across all industries, providing crowd-sourced expertise and feedback. 

Based on these collective insights, we describe the design decisions and trade-offs that lead to the development of the next version of EPSS. This new machine learning model provides an 82\% performance improvement over past models in distinguishing vulnerabilities that are exploited in the wild and thus may be prioritized for remediation.

\end{abstract}

\section{Introduction}

Vulnerability management, the practice of identifying, prioritizing, and patching known software vulnerabilities, has been a continuous challenge for defenders for decades. This issue is exacerbated by the increasing number of new vulnerabilities that are being disclosed annually. For example, MITRE published\footnote{Not marked as REJECT or RESERVED.} 25,068 new vulnerabilities during the 2022 calendar year, a 24.3\% increase over 2021.

Adding to the increasing rate of published vulnerabilities are challenges incurred by practitioners when trying to remediate them. Recent research conducted by Kenna Security and Cyentia tracked exposed vulnerabilities at hundreds of companies and found that the monthly median rate of remediation was only 15.5\%, while a quarter of companies remediated less than 6.6\% of their open vulnerabilities per month~\cite{cyentia2022p2pv8}. As a consequence of the increasing awareness of software flaws and the limited capacity to remediate them, vulnerability prioritization has become both a chronic and an acute concern for every organization attempting to reduce their attack surface. 

The prioritization process involves scoring and ranking vulnerabilities according to assessments, often based on the industry standard Common Vulnerability Scoring System (CVSS)~\cite{cvss3guide}. However, only the Base metric group of CVSS is being assigned and distributed at scale by NIST, and this group of metrics is unable to adapt to post-disclosure information, such as the publication of exploits or technical artifacts, which can affect the odds of attacks against a vulnerability being observed in the wild.
As a result, while only 5\% of known vulnerabilities are exploited in the wild~\cite{jacobs2020improving}, numerous prior studies have shown that CVSS does not perform well when used to prioritize exploited vulnerabilities over those without evidence of exploitation~\cite{Allodi12:VulnerabilityScores,eiram2013exploitability,allodi2014comparing}.
While several other efforts have been made to capture exploitation likelihood in vulnerability assessments, these approaches are either vendor-specific~\cite{MS:ExploitabilityIndex,RedHat:SeverityRating} or proprietary and not available publicly~\cite{VPRDoc,Rapid7Risk,recordedfuturerisk}. 

\noindent\fbox{%
    \parbox{\linewidth}{%
In order to improve remediation practices, network defenders need a scoring systems that can accurately quantify \emph{likelihood of exploits in the wild}, and is able to \emph{adapt to new information} published after the initial disclosure of a vulnerability.
    }
}

Any effort to developing a new capability to understand, anticipate, and respond to new cyber threats must overcome three main challenges: i) it must address the requirements of practitioners who rely on it; ii) it must provide significant performance improvements over existing scoring systems; and iii) it must have a low barrier for adoption and use. 

To address these challenges, a Special Interest Group (SIG) was formed in early 2020 at the Forum of Incident Response and Security Teams (FIRST). From its inception, the Exploit Prediction Scoring System (EPSS) SIG has gathered 170 members from across the world, representing practitioners, researchers, government agencies, and software developers.\footnote{See \url{https://www.first.org/epss}.}
The SIG was created with the publication of the first EPSS model for predicting the likelihood of exploits in the wild~\cite{jacobs2021epss} and is organized around a mailing list, a discussion forum, and bi-weekly meetings. 
This unique environment represented an opportunity to understand the challenges faced by practitioners when performing vulnerability prioritization, and therefore address the first challenge raised above by designing a scoring system that takes into account practitioner requirements.

To address the second challenge and achieve significant performance improvements, the SIG provided subject matter expertise, which guided feature engineering with high utility at predicting exploits in the wild. Finally, to address the challenges of designing a public and readily-available scoring system, the SIG attracted a set of industry partners willing to share proprietary data for the development of the model, the output of which can then be made public. This allowed EPSS scores to be publicly available at scale,  lowering the barrier to entry for those wanting to integrate EPSS into their prioritization pipeline.   

This paper presents the latest (third) iteration of the EPSS model, as well as lessons learned in its design, and their impact on designing a scoring system. The use of a novel and diverse feature set and optimized machine learning techniques allows EPSS to improve prediction performance by 82\% over its predecessor (as measured by the precision/recall Area Under the Curve improved to 0.779 from 0.429). EPSS is able to score all vulnerabilities published on MITRE's CVE List (and the National Vulnerability Database), and can reduce the amount of effort required to patch critical vulnerabilities to one-eighth of a comparable strategy based on CVSS.
This paper makes the following contributions:
\begin{enumerate}
    \item Present lessons learned from developing an exploit prediction model that integrates the functional requirements of a community of nearly 200 practitioners and researchers.
    \item Engineers novel features for exploit prediction and use them to train the EPSS classifier for predicting the likelihood of exploits in the wild. 
    \item Analyzes the practical utility of EPSS by showing that it can significantly improve remediation strategies compared to static baselines. 
\end{enumerate}
\section{Evolution of EPSS}

EPSS was initially inspired by the Common Vulnerability Scoring System (CVSS). The first EPSS model~\cite{jacobs2021epss} was designed to be lightweight, portable (i.e. implemented in a spreadsheet), and parsimonious in terms of the data required to score vulnerabilities. Because of these design goals, the first model used a logistic regression which produced interpretable and intuitive scores, and predicted the probability of exploitation activity being observed in the first year following the publication of a vulnerability. In order to be parsimonious, the logistic regression model was trained on only 16 independent variables (features) extracted at the time of vulnerability disclosure. While outperforming CVSS, the SIG highlighted some key limitations which hindered its practical adoption.

Informed by this feedback, the second version of EPSS aimed to address the major limitations of the first version. The first design decision was to switch to a centralized architecture. By centralizing and automating the data collection and scoring, a more complex model could be developed to improve performance. This decision came with a trade-off, namely a  loss of the model's portability and thus, the ability to score vulnerabilities which are not publicly disclosed (e.g., zero day vulnerabilities, or flaws that may never be assigned a CVE ID). Nevertheless, focusing on public vulnerabilities under the centralized model removed the need for each implementation of EPSS to perform their own data collection, and further allowed more complex features and models. The model used in v2 is XGBoost~\cite{chen2016xgboost}, and the feature set was greatly expanded from 16 to 1,164. These efforts led to a significant improvement in predictive performance over the previous version by capturing higher order interactions in the extended feature set. Another major component of a centralized architecture was being able to adapt to new vulnerability artifacts (e.g., the publication of exploits) and produce new predictions, daily. Moreover, the SIG also commented that producing scores based on the likelihood of exploitation within the first year of a vulnerability's lifecycle was not very practical, since most prioritization decisions are made with respect to an upcoming patching cycle. As a result, v2 switched to predicting exploitation activity within the following 30-day window as of the time of scoring, which aligns with the typical remediation window of practitioners in the SIG.


For the third version of EPSS, the SIG highlighted a requirement for improved precision at identifying vulnerabilities likely to be exploited in the wild. 
This drove an effort to expand the sources of exploit data by partnering with multiple organizations willing to share data for model development, and engineer more complex and informative features. 
These label and feature improvements, along with a methodical hyper-parameter tuning approach, enabled improved training of an XGBoost classifier. This allowed the proposed v3 model to achieve an overall 82\% improvement in classifier performance over v2, with the Area Under the Precision/Recall Curve increasing from 0.429 to 0.779. 
This boost in prediction performance allows organizations to substantially improve their prioritization practices and design data-driven patching strategies.

\section{Data}

The data used in this research is based on 192,035 published vulnerabilities (not marked as ``REJECT'' or ``RESERVED'') listed in MITRE's Common Vulnerabilities and Exposures (CVE) list through December 31, 2022. The CVE identifier has been used to combine records across our disparate data sources. Table \ref{table:features} lists the categories of data, number of features in each category, and the source(s) or other notes. In total, EPSS collects 1,477 unique independent variables for every vulnerability.

\begin{table*}[t]
    \caption{Description of data sources used in EPSS.}
    \label{table:features}
    \begin{tabular}{p{0.35\linewidth} p{0.1\linewidth} p{0.1\linewidth} p{0.35\linewidth}}
        \hline
        Description & \# of variables & Type & Sources \\
        \hline
        Exploitation activity in the wild (labels) & 1 (with dates) & Binary & Fortinet, AlienVault, Shadowserver, GreyNoise \\
        Publicly available exploit code & 3 & Binary & Exploit-DB, GitHub, MetaSploit \\
        CVE mentioned on list or website & 3 & Binary & CISA KEV, Google Project Zero, Trend Micro ZDI \\
        Social media & 3 & Numeric & Mentions/discussion on Twitter \\
        Offensive security tools and scanners & 4 & Binary & Intrigue, sn1per, jaeles, nuclei \\
        References with labels & 17 & Numeric & MITRE CVE List, NVD \\
        Keyword description of vulnerability & 147 & Binary & Text description in MITRE CVE List \\
        CVSS metrics & 15 & One-Hot & National Vulnerability Database (NVD) \\
        CWE & 188 & Binary & National Vulnerability Database (NVD) \\
        Vendor labels & 1,096 & Binary & National Vulnerability Database (NVD) \\
        Age of the vulnerability & 1 & Numeric & Days since CVE published in MITRE CVE list \\
        \hline
    \end{tabular}
\end{table*}

\subsection{Labeling data: exploitation in the wild}

EPSS collects and aggregates evidence of exploits from multiple sources: Fortiguard, Alienvault OTX, the Shadowserver Foundation and GreyNoise (though not all sources cover the full time period). Each of these data sources employ network- or host-layer intrusion detection/prevention systems (IDS/IPS), or honeypots, in order to identify attempted exploitation. These systems are also predominantly signature-based (as opposed to anomaly-based) detection systems. Moreover, all of these organizations have large enterprise infrastructures of sensor and collection networks. Fortiguard, for example, manages tens of thousands of IDS/IPS devices that identify and report exploitation activity from across the globe. Alienvault OTX, GreyNoise and the Shadowserver Foundation also maintain worldwide networks of sensors for detecting exploitation activity. Aggregating exploit evidence from multiple sources does not guarantee uniform coverage of labels across all types of vulnerabilities, and this could lead to class- and feature-dependent noise when used to train machine learning models~\cite{suciu2022expected}. We discuss these limitations in Section~\ref{section:discussion}.

These data sources include the list of CVEs observed to be exploited on a daily basis. The data are then cleaned, and exploitation activity is consolidated into a single boolean value (0 or 1), identifying days on which exploitation activity was reported for any given CVE across any of the available data sources. Structuring the training data according to this boolean time-series enables us to estimate the probability of exploitation activity in any upcoming window of time, though the consensus in the EPSS Special Interest Group was to standardize on a 30-day window to align with most enterprise patch cycles. 

\noindent\fbox{%
    \parbox{\linewidth}{%
The exploit data used in this research paper cover activity from July 1, 2016 to December 31st, 2022 (2,374 days / 78 months / 6.5 years), over which we collected 6.4 million exploitation observations (date and CVE combinations), targeting 12,243 unique vulnerabilities. Based on these data, we find that 6.4\% (12,243 of 192,035) of all published vulnerabilities were observed to be exploited during this period, which is consistent with previous findings \cite{jacobs2020improving, jacobs2021epss}.
    }%
}

\subsection{Explanatory variables/features}
\label{subsec:explan_variables}

In total, EPSS leverages 1,477 features for predicting exploitation activity. Next, we describe the data sources  used to construct these features as well as the engineering behind them. 

\paragraph{Published exploit code}

We first consider the correlation between exploitation in the wild and the existence of publicly available exploit code, which is collected from three sources (courtesy of Cyentia\footnote{\url{https://www.cyentia.com/services/exploit-intelligence-service}}): Exploit-DB, Github, and Metasploit. In total we identified 24,133 CVEs with published exploit code, consisting of 20,604 CVEs from Exploit-DB, 4,049 published on GitHub, and 1,905 published on Metasploit modules. Even though Exploit-DB contains the majority of published exploits, GitHub has become a valuable source in recent years. For example, in 2022, 1,591 exploits were published on GitHub, while Exploit-DB and Metasploit added 196 and 94 entries, respectively. We derive three binary features from this category.

\paragraph{Public vulnerability lists}

Next, we consider that exploitation activity may be forecasted by the presence of vulnerabilities on popular lists and/or websites that maintain and share information about selective vulnerabilities.  Google Project Zero maintains a listing\footnote{\url{https://docs.google.com/spreadsheets/d/1lkNJ0uQwbeC1ZTRrxdtuPLCIl7mlUreoKfSIgajnSyY/view\#gid=1190662839}.} of ``publicly known cases of detected zero-day exploits.''\footnote{\url{https://googleprojectzero.blogspot.com/p/0day.html}.} This may help forecast exploitation activity as the vulnerability slides into N-day status. We include 162 unique CVEs listed by Google Project Zero.

Trend Micro's Zero Day Initiative (ZDI), the ``world's largest vendor-agnostic bug bounty program'',\footnote{\url{https://www.zerodayinitiative.com/about}.} works with researchers and vendors to responsibly disclose zero-day vulnerabilities and issue public advisories about vulnerabilities at the conclusion of their process. We include 7,356 CVEs that have public advisories issued by ZDI. 

The Known Exploited Vulnerabilities (KEV) catalog from the US Department of Homeland Security's Cybersecurity and Infrastructure Security Agency (CISA) is an ``authoritative source of vulnerabilities that have been exploited in the wild''.\footnote{\url{https://www.cisa.gov/known-exploited-vulnerabilities}} We include 866 CVEs from CISA's KEV list.

These sources lack transparency about when exploitation activity was observed, and for how long this activity was ongoing. However, because past exploitation attempts might influence the likelihood of future attacks, we include these indicators as binary features for our model.

\paragraph{Social media}

Exploitation may also be correlated with social media discussions, and therefore we collect Twitter mentions of CVEs, creating three features counting these mentions within three different historical time windows (7, 30, and 90 days). We only count primary and original tweets and exclude retweets and quoted retweets. The median number of daily unique tweets mentioning CVEs is 1,308 with the 25th and 75th percentile of daily tweets being 607 and 1,400 respectively. We currently make no attempt to validate the content or filter out automated posts (from bots). 

\paragraph{Offensive security tools}

We also collect evidence of vulnerabilities being used in offensive security tools that are designed, in part, to identify vulnerabilities during penetration tests. We are currently gathering information from four different offensive security tools with varying numbers of CVEs identified in each: Nuclei with 1,548 CVEs, Jaeles with 206 CVEs, Intrigue with 169 CVEs and Sn1per with 63 CVEs. These are encoded as binary features which indicate whether each particular source is capable of scanning for and reporting on the presence of each vulnerability. 

\paragraph{References}

In order to capture metrics around the activity and analysis related to vulnerabilities, for each CVE, we count the number of references listed in MITRE's CVE list, as well as the number of references with each of the 16 reference tags assigned by NVD. The labels and their associated prevalence across CVEs are: Vendor Advisory (102,965), Third Party Advisory (84,224), Patch (59,660), Exploit (54,633), VDB Entry (31,880), Issue Tracking (16,848), Mailing List (15,228), US Government Resource (11,164), Release Notes (9,308), Permissions Required (3,980), Broken Link (3,934), Product (3,532), Mitigation (2,983), Technical Description (1,686), Not Applicable (961), and Press/Media Coverage (124).

\paragraph{Keyword description of the vulnerability}

To capture attributes of vulnerabilities themselves, we use the same process as described in previous research~\cite{jacobs2020improving, jacobs2021epss}. This process detects and extracts hundreds of common multiword expressions used to describe and discuss vulnerabilities. These expressions are then grouped and normalized into common vulnerability concepts. The top tags we included and associated CVEs are as follows: ``remote attacker'' (80,942), ``web'' (31,866), ``code execution'' (31,330), ``denial of service'' (28,478), and `authenticated'' (21,492). In total, we include 147 binary features for identifying such tags. 

We followed the same process as EPSS v1 for extracting multi-word expressions from the text from references using Rapid Automatic Keyword Extraction~\cite{rose2010automatic}.

   
\paragraph{CVSS metrics}

To capture other attributes of vulnerabilities, we collect CVSS base metrics. These consist of exploitability measurements (attack vector, attack complexity, privilege required, user interaction, scope) and the three impact measurements (confidentiality, integrity and availability). These categorical variables are encoded using one-hot encoding. We collected CVSS version 3 information from NVD for 118,087 vulnerabilities. However, 73,327 vulnerabilities published before CVSSv3 were created and are only scored in NVD using CVSSv2. To address this, we developed a separate and dedicated machine learning model to estimate the CVSSv3 measurement values for each of these vulnerabilities. 

We use a process similar to prior work~\cite{nowak2021conversion}, where for each CVE, we use the CVSSv2 sub-components for CVEs which have both CVSSv2 and CVSSv3 scores. We then train a feedforward neural network to predict CVSSv3 vectors. The model was validated using 8-fold, yearly stratified, cross-validation, achieving 74.9\% accuracy when predicting the exact CVSSv3 vector. For 99.9\% of vectors, we predict the majority (5 or more) of the individual metrics correctly. For each individual portion of the CVSSv3 vector we were able to achieve a minimum of 93.4\% accuracy (on the Privileges Required metric). We note that this exceeds the accuracy achieved by \cite{nowak2021conversion}, and likely warrants further research into the robustness of CVSSv3 prediction and its possible application to future versions of CVSS.

\paragraph{CWE}

We also capture the observation that different types of vulnerabilities may be more or less attractive to attackers, using the Common Weakness Enumeration (CWE), which is a ``community-developed list of software and hardware weakness types.''\footnote{\url{https://cwe.mitre.org}} We collect the CWE assignments from NVD, noting that 21,570 CVEs do not have a CWE assigned. We derived binary features for CWEs found across at least 10 vulnerabilities, resulting in 186 CWE identifiers being included. In addition, we maintain two features for vulnerabilities where CWE information is not available, or the assigned CWEs are not among the common ones. The top CWE identifiers and their vulnerability counts are CWE 79 (20,797), CWE 119 (11,727), CWE 20 (9,590), CWE 89 (8,790), CWE 787 (7,624), CWE 200 (7,270), CWE 264 (5,485), CWE 22 (4,918), CWE 125 (4,743), and CWE 352 (4,081).

\paragraph{Vulnerable vendors}

We suspect exploitation activity may be correlated to the market share and/or install base companies achieve. Therefore, we parse through the Common Platform Enumeration (CPE) data provided by NVD in order to identify platform records marked as ``vulnerable'', and extract only the vendor portion of the record. We did not make any attempt to fill in missing information or correct any typos or misspellings that may occasionally appear in the records. We ranked vendors according to the number of vulnerabilities, creating one binary feature for each vendor, and evaluated the effect of including less frequent vendors as features. We observed no performance improvements by including vendors with fewer than 10 CVEs in our dataset. As a result, we extracted 1,040 unique vendor features in the final model. The most prevalent vendors and their vulnerability counts are Microsoft (10,127), Google (9,100), Oracle (8,970), Debian (7,627), Apple (6,499), IBM (6,409), Cisco (5,766), RedHat (4,789), Adobe (4,627), Fedora Project (4,166). 

\paragraph{Age of the vulnerability}

Finally, the age of a vulnerability might contribute or detract from the likelihood of exploitation. Intuitively, we expect old vulnerabilities to be less attractive to attackers due to a smaller vulnerable population. To capture this, we create a feature which records the number of days elapsed from CVE publication to the time of feature extraction in our model. 

\section{Modeling Approach}

\subsection{Preparing labels and features}

Exploitation activity is considered as any recorded attempt to exploit a vulnerability, regardless of the success of the attempt, and regardless of whether the targeted vulnerability is present. All observed exploitation activity is recorded with the date the activity occurred and aggregated across all data sources by the date and CVE identifier. The resulting labeling data is a binary value for each vulnerability of whether exploitation activity was observed or not, for each day.

Since many of the features may change day by day, we construct features for the training data on a daily basis. In order to reduce the size of our data (and thus the time and memory needed to train models) we aggregate consecutive daily observations where features do not change.  The size of the exposure and the number of days with exploitation activity are included in the model training. 

When constructing the test data, a single date is selected (typically "today", see next section) and all of the features are generated based on the state of vulnerabilities for that date. Since the final model is intended to estimate the probability of exploitation in the next 30 days, we construct labels for the test data by looking for exploitation activity over the following 30 days from the test date selected.

\subsection{Model selection}

The first EPSS model~\cite{jacobs2021epss} sought not only to accurately predict exploitation but do so in a parsimonious, easy to implement way. As a result, regularized logistic regression (Elasticnet) was chosen to produce a generalized linear model with only a handful of variables. The current model relaxes this requirement in the hopes of improving performance and providing more accurate exploitation predictions. In particular, capturing non-linear relationships between inputs and exploitation activity will better predict the finer exploitation activity.

Removing the requirement of a simple model with the need to model complex relationships expands the universe of potential models. Indeed many machine learning algorithms have been developed for this exact purpose. However, testing all models is impractical because each model requires significant engineering and calibration to achieve an optimal outcome. We therefore focus on a single type of model that has proven to be particularly performant on these data. Recent research has illustrated that panel (tabular) data, such as ours, can be most successfully modeled using tree based methods (in particular gradient boosted trees for regression)~\cite{grinsztajn2022tree}, arriving at similar or better predictive performance with less computation and tuning in comparison to other methods such as neural networks. Given the results in \cite{grinsztajn2022tree} we focus our efforts on tuning a common implementation of gradient boosted trees~\cite{chen2016xgboost}. We also provide a comparison to a transformer-based neural network in \autoref{section:neural-net}.

\noindent\fbox{%
    \parbox{\linewidth}{%
XGBoost is a popular, well documented, and performant implementation of the gradient boosted tree algorithm in which successive decision trees are trained to iteratively reduce prediction error. 
    }%
}

\subsection{Train/test split and measuring performance}

In order to reduce overfitting, we implement two restrictions. First, we implement a time-based test/train split, constructing our training data sets on data up to and including October 31, 2021. We then construct the test data set based on the state of vulnerabilities on December 1st, 2021, providing one month between the end of the training data and the test data. As mentioned above, the ground truth in the test data is any exploitation activity from December 1st to December 30th, 2021. Second, we use 5-fold cross validation, with the folds based on each unique CVE identifier. This selectively removes vulnerabilities from the training data and tests the performance on the hold out set, thus further reducing the likelihood of overfitting. We chose $k=5$ for our procedure as it corresponds to an 80\%/20\% split in training and test data. This larger validation size (20\%) is less likely to induce overfitting, and therefor poor hyper parameter selection, than a $k=10$ (90\%/10\% train/test split) as described in~\cite{cawley2010over}.  Additionally, we stratify the folds to ensure the same proportion of exploitation activity for each fold as recommended in~\cite{kohavi1995study}. Other values of $k$ may provide better performance, but due to computational restraints we rely on the literature as a guide for this particular parameter rather than adding an additional dimension to our model search space.

Finally, we measure performance by calculating the area under the curve (AUC) based on precision and recall across the full range of predictions. We selected precision-recall since we have severe class imbalance in exploited vulnerabilities, and using accuracy or traditional Receiver Operator Characteristic (ROC) curves may be misleading due to that imbalance. 

\subsection{Tuning and optimizing model performance}

Despite being a well studied approach, the use of gradient boosted trees and XGBoost for prediction problems still requires some effort to identify useful features and model tuning to achieve good model performance. This requires a-priori decisions about which features to include and the hyperparameter values for the XGBoost algorithm.

The features outlined in \autoref{subsec:explan_variables} includes 28,724 variables. Many of these variables are binary features indicating whether a vulnerability affects a particular vendor or can be described by a specific CWE. While the XGBoost algorithm is efficient, including all variables in our inference is technically infeasible. To reduce the scope of features we take a naive, yet demonstrably effective approach at removing variables below a specific occurrence rate~\cite{yang1997comparative}. This reduced the input feature set to 1,477 variables.

One additional challenge with our data is the temporal nature of our predictions. In particular, exactly how much historical data should be included in the data set. In addition to the XGBoost hyperparameters and the sparsity threshold, we also constructed four different sets of training data for 6 months and then 1, 2 and 3 years, to determine what time horizons would provide the best predictions.

To identify the time horizon and sparsity threshold described above as well as the other hyperparameters needed by our implementation of gradient boosted trees we take a standard approach described in \cite{yang2020hyperparameter}. We first define reasonable ranges for the hyperparameters, use Latin Hypercube sampling over the set of possible combinations, compute model performance for that set of hyperparameters, then finally build an additional model (also a gradient boosted tree) to predict performance given a set of hyperparameters, using the model to maximize performance.

\begin{table}[t]
    \caption{Non-default hyperparameter values for XGBoost algorithm and data selection}
    \label{table:hyperparams}
    \begin{tabular}{ | m{20em} | r | }
    \hline
    \textbf{Parameter} & \textbf{Value} \\
    \hline
        Time Horizon & 1 year \\ 
    \hline
        Learning rate & 0.11 \\ 
    \hline
        Max depth tree depth & 20 \\
    \hline
        Subsample ratio of the training instances & 0.75 \\
    \hline
        Minimum loss reduction for leaf node partition & 10 \\
    \hline
        Maximum delta step & 0.9 \\
    \hline
        The number of boosting rounds & 65 \\
    \hline
    \end{tabular}
\end{table}

The results of the above process results in the parameters selected in Table \ref{table:hyperparams}. Note that of the tested time horizons, none dramatically outperformed others, with 1 year only slightly outperforming other tested possibilities.

\section{Evaluation}\label{section:evaluation}

\subsection{Precision (efficiency) and recall (coverage)}

Precision and recall are commonly used machine learning performance metrics, but are not intuitive for security practitioners, and therefore can be difficult to contextualize what these performance metrics represent in practice.

Precision (efficiency) measures how well resources are being allocated, (where low efficiency represents wasted effort), and is calculated as the true positives divided by the sum of the true and false positives.

\noindent\fbox{%
    \parbox{\linewidth}{%
In the vulnerability management context, efficiency addresses the question, ``out of all the vulnerabilities remediated, how many were actually exploited?'' If a remediation strategy suggests patching 100 vulnerabilities, 60 of which were exploited, the efficiency would be 60\%. 
    }%
}
Recall (coverage), on the other hand, considers how well a remediation strategy actually addresses those vulnerabilities that should be patched (e.g., that have observed exploitation activity), and is calculated as the true positives divided by the sum of the true positives and false negatives.

\noindent\fbox{%
    \parbox{\linewidth}{%
In the vulnerability management context, coverage addresses the question, ``out of all the vulnerabilities that are being exploited, how many were actually remediated?'' If 100 vulnerabilities are exploited, 40 of which are patched, the coverage would be 40\%. 
    }%
}
Therefore, for the purpose of this article, we use the terms efficiency and coverage interchangeably with precision and recall, respectively, in the discussions below.

\subsection{Model performance}

\begin{figure}[t]
	\centering
	\includegraphics[width=1\columnwidth]{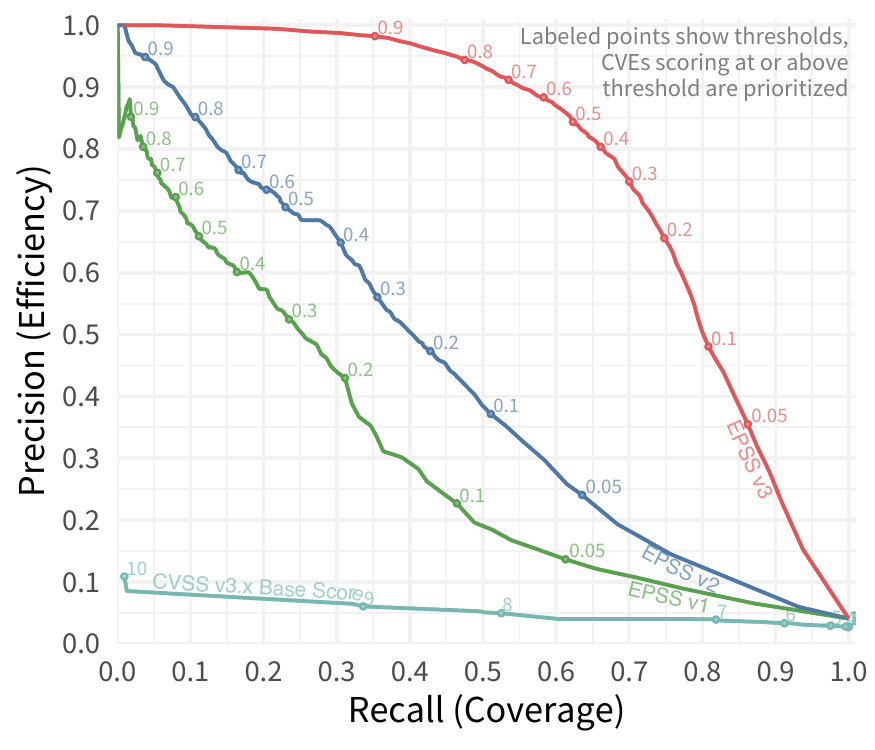}
	\caption{Performance of EPSS v3 compared to previous versions and CVSS Base Score}
	\label{figure:precision-recall}
\end{figure}

After several rounds of experiments to find the optimal set of features, amount of historical data, and model parameters as discussed in the previous section, we generated one final model using all vulnerabilities from November 1st, 2021 to October 31st, 2022. We then predicted the probability of exploitation activity in the next 30 days based on the state of vulnerabilities on December 1st, 2022. Using evidence of exploitation activity for the following 30 days (through Dec 30th, 2022), we measured  overall performance as shown in Figure \ref{figure:precision-recall}. For comparison, we also show performance metrics for the EPSS versions 1 and 2, as well as CVSS v3 base scores for the same date and exploitation activity (Dec 1st, 2022). Figure \ref{figure:precision-recall} includes points along the precision-recall curves that represent the thresholds with each prioritization strategy. 

Figure \ref{figure:precision-recall} clearly illustrates the significant improvement of the EPSS v3 model over previous versions, as well as the CVSS version 3 base score. 

\noindent\fbox{%
    \parbox{\linewidth}{%
EPSS v3 produces an area under the curve (AUC) of 0.7795, and an F1 score of 0.728. A remediation strategy based on this F1 score would prioritize remediation for vulnerabilities with EPSS probabilities of 0.36 and above, and would achieve an efficiency of 78.5\% and coverage of 67.8\%. 
    }%
}
In addition, this strategy would prioritize remediation of 3.5\% of all published vulnerabilities (representing the level of effort).

EPSS v2 has an AUC of 0.4288 and a calculated F1 score at 0.451, which prioritizes vulnerabilities with a probability of 0.16 and above. At the F1 threshold, EPSS v2 achieves an efficiency rating of 45.5\% and coverage of 44.8\% and prioritizes 4\% of the vulnerabilities in our study. EPSS v1 has an AUC of 0.2998 and a calculated F1 score at 0.361, which prioritizes vulnerabilities with a probability of 0.2 and above. At the F1 threshold, EPSS v1 achieves an efficiency rating of 43\% and coverage of 31.1\% and prioritizes 2.9\% of the vulnerabilities in our study.  Finally, CVSS v3.x base score has an AUC of 0.051 and a calculated F1 score at 0.108, which prioritizes vulnerabilities with a CVSS base score of 9.7 or higher. At the F1 threshold, CVSS v3.x achieves an efficiency rating of 6.5\% and coverage of 32.3\% and prioritizes 13.7\% of the vulnerabilities in our study.

\subsection{Probability calibrations}

A significant benefit of this model over alternative exploit scoring systems (described above) is that the output scores are true  probabilities (i.e., probability of any exploitation activity being observed in the next 30 days) and can therefore be scaled  to produce a threat score based on one or more vulnerabilities, such as would be found in a single network device (laptop, server), network segment, or an entire enterprise. For example, standard mathematical techniques can be used to answer questions like ``what is the probability that at least one of this asset's vulnerabilities will be exploited in the next 30 days?'' Such estimates, however, are only useful if they are calibrated and therefore reflect the true likelihood of the event occurring. 

\begin{figure}[t]
	\centering
	\includegraphics[width=1\columnwidth]{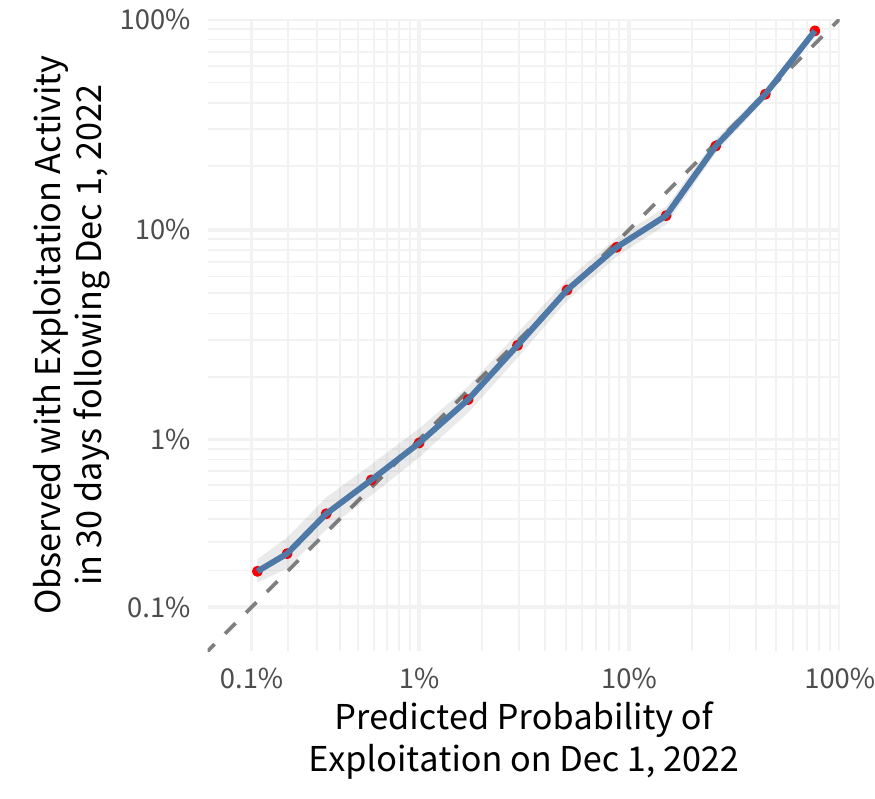}
	\caption{Calibration Plot comparing predicted probabilities to observed exploitation period in the following 30 days}
	\label{figure:calibration}
\end{figure}

In order to address this, we measure calibration in a two ways. First we calculate a Brier Score~\cite{brier1950verification} which produces a score between 0 and 1, with 0 being perfectly calibrated and 1 being perfectly uncalibrated (the original 1950 paper doubles the range from 0 to 2). Our final estimate revealed a Brier score of 0.0162, which is objectively very low (good). We also plot the predicted (binned) values against the observed (binned) exploitation activity (commonly referred to as a ``calibration plot'') as shown in Figure \ref{figure:calibration}. The closer the plotted line is to a 45 degree line (i.e. a line with a slope of 1, represented by the dashed line), the greater the calibration. Again, by visual inspection, our plotted line very closely matches the 45 degree line. 

\subsection{Simple Remediation Strategies}
Research conducted by Kenna Security and Cyentia tracked vulnerabilities at hundreds of companies and found that on average, companies were only able to remediate about 15.5\% of their open vulnerabilities in a month\cite{cyentia2022p2pv8}. This research also found that resource capacity for remediating vulnerabilities varies considerably across companies, which suggests that any vulnerability remediation strategy should accommodate varying levels of corporate resources and budgets. Indeed, organizations with fewer resources (presumably smaller organizations) may prefer to emphasize efficiency over coverage, to optimize their spending, while larger organizations may accept less efficient strategies in exchange for the greater coverage (i.e. more vulnerabilities patched). 

\begin{figure}[t]
   \centering
   \includegraphics[width=1\columnwidth]{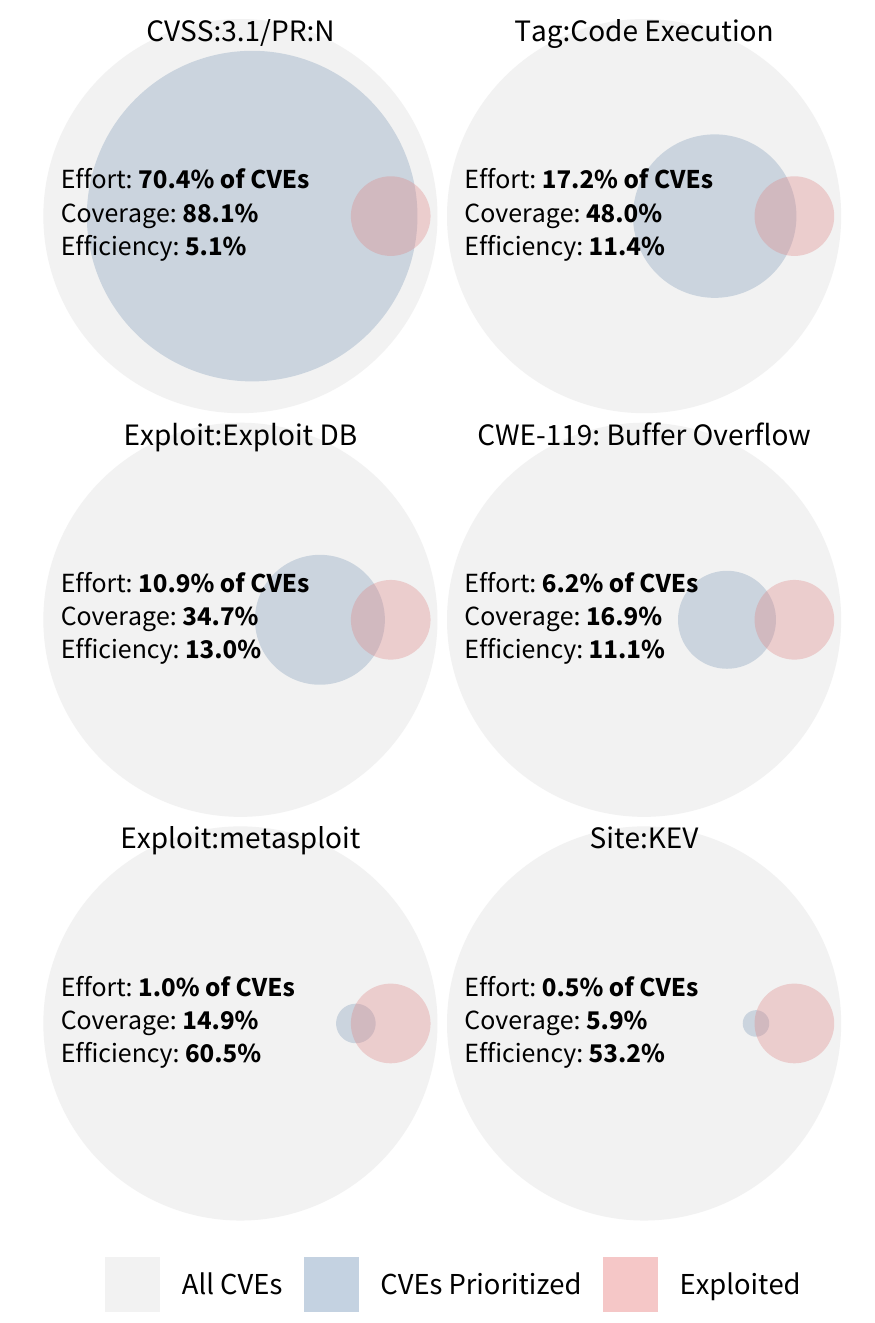}
   \caption{Alternative strategies based on simple heuristics}
   \label{figure:alternates}
\end{figure}

Therefore, we compare the amount of effort required (as measured by the number of vulnerabilities needing to be remediated) for differing remediation strategies. Figure \ref{figure:alternates} highlights the performance of 6 simple (but practical) vulnerability prioritization strategies based on our test data (December 1st, 2022).\footnote{Performance is then measured based on exploitation activity in the following 30 days.}

The first diagram in the upper row considers a strategy based on the CVSS v3.x vector of ``Privilege Required: None''. Being able to exploit a vulnerability that doesn't require any established account credentials is an attractive vulnerability to exploit, as an attacker. 
\noindent\fbox{%
    \parbox{\linewidth}{%
While this strategy would yield 88.1\% coverage, it would achieve only 5.1\% efficiency. That is, from a defender perspective, this class of vulnerabilities represents over 130,000 (70\%) of all published CVEs, and would easily surpass the resources capacity of most organizations.
    }%
}
``Code Execution'' is another attractive vulnerability attribute for attackers since these vulnerabilities could allow the attacker to achieve full control of a target asset. However, remediating all the code execution vulnerabilities (17\% or about 32,000 of all CVEs) would achieve 48\% coverage and 11.4\% efficiency.

The middle row of Figure \ref{figure:alternates} shows remediation strategies for vulnerabilities published in Exploit DB (left), and Buffer Overflows (CWE-119; right3), respectively. 

The bottom row of Figure \ref{figure:alternates} is especially revealing. The bottom right diagram shows performance metrics for a remediation strategy based on patching vulnerabilities from the Known Exploited Vulnerabilities (KEV) list (as of Dec 1, 2022) from DHS/CISA. The KEV list is meant to prioritize vulnerability remediation for US Federal agencies as per Binding Operational Directive 22-01\footnote{"See https://www.cisa.gov/binding-operational-directive-22-01"}. Strictly following the KEV would remediate half of one percent (0.5\%) of all published CVEs, and produce a relatively high efficiency of 53.2\%. However, with almost 8,000 unique CVEs with exploitation activity in December, the coverage obtained from this strategy is only 5.9\%. 

Alternatively, the strategy identified in the bottom left diagram  shows a remediation strategy based on whether a vulnerability appears in a Metasploit module. In this case, a network defender would need to remediate almost twice as many vulnerabilities on the KEV list, but would enjoy 13\% greater efficiency (60.5\% vs 53.2\%) and almost three times more coverage (14.9\% vs 5.9\%). 

\noindent\fbox{%
    \parbox{\linewidth}{%
Therefore, based on this simple heuristic (KEV vs Metasploit), the Metasploit strategy outperforms the KEV strategy.
    }%
}
\subsection{Advanced remediation strategies}

Next we explore the real-world performance of our model, using two separate approaches. We first compare coverage among four remediation strategies while holding the \textit{level of effort} constant (i.e. the number of vulnerabilities needing to be remediated), we then compare levels of effort while holding  \textit{coverage} constant. 

\begin{figure}[t]
   \centering
   \includegraphics[width=1\columnwidth]{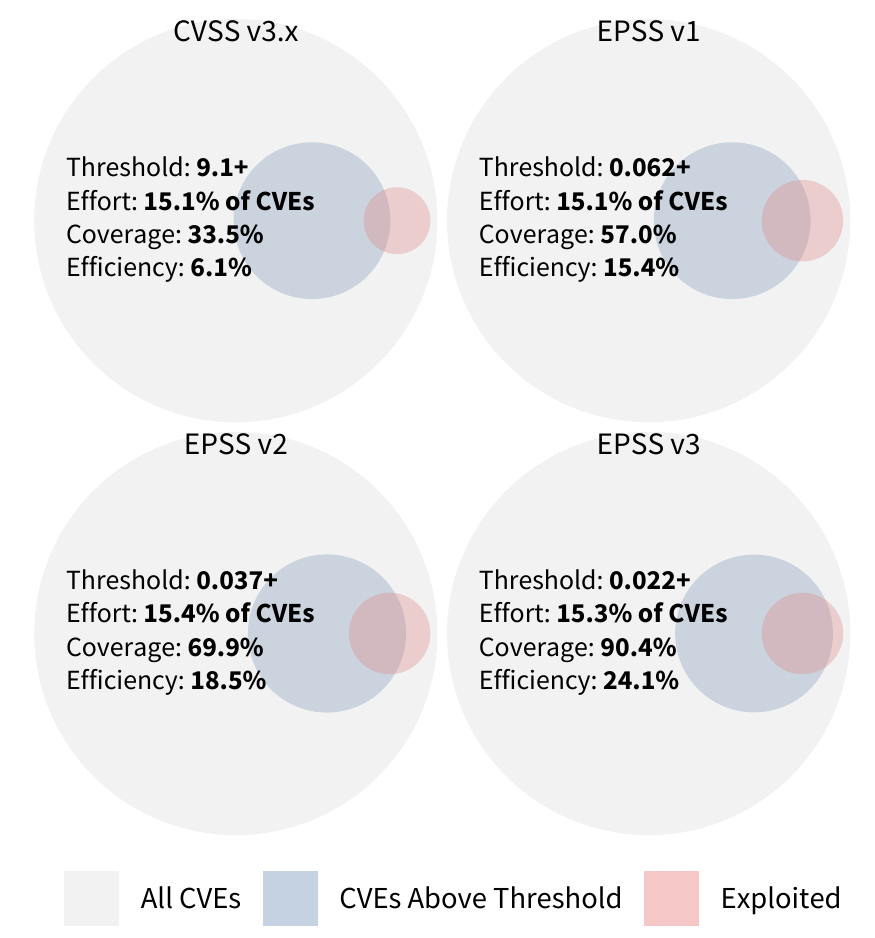}
   \caption{Strategy comparisons holding the level of effort constant}
   \label{figure:venn_holding_effort}
\end{figure}

Figure \ref{figure:venn_holding_effort} compares the four strategies while maintaining approximately the same level of effort. That is, the blue circle in the middle of each figure -- representing the number of vulnerabilities that would need to be remediated -- is fixed to the same size for each strategy, at approximately 15\% or about 28,000 vulnerabilities. The CVSS strategy, for example, would remediate vulnerabilities with a base score of 9.1 or greater, and would achieve coverage and efficiency of 33.5\% and 6.1\%, respectively. 

A remediation strategy based on EPSS v2, on the other hand, would remediate vulnerabilities with an EPSS v2 score of 0.037 and greater, yielding 69.9\% coverage and 18.5\% efficiency. Already, this strategy doubles the coverage and triples the efficiency, relative to the CVSS strategy. 

Even better results are achieved with a remediation strategy based on EPSS v3 which enjoys 90.4\% coverage and 24.1\% efficiency.

\begin{figure}[t]
   \centering
   \includegraphics[width=1\columnwidth]{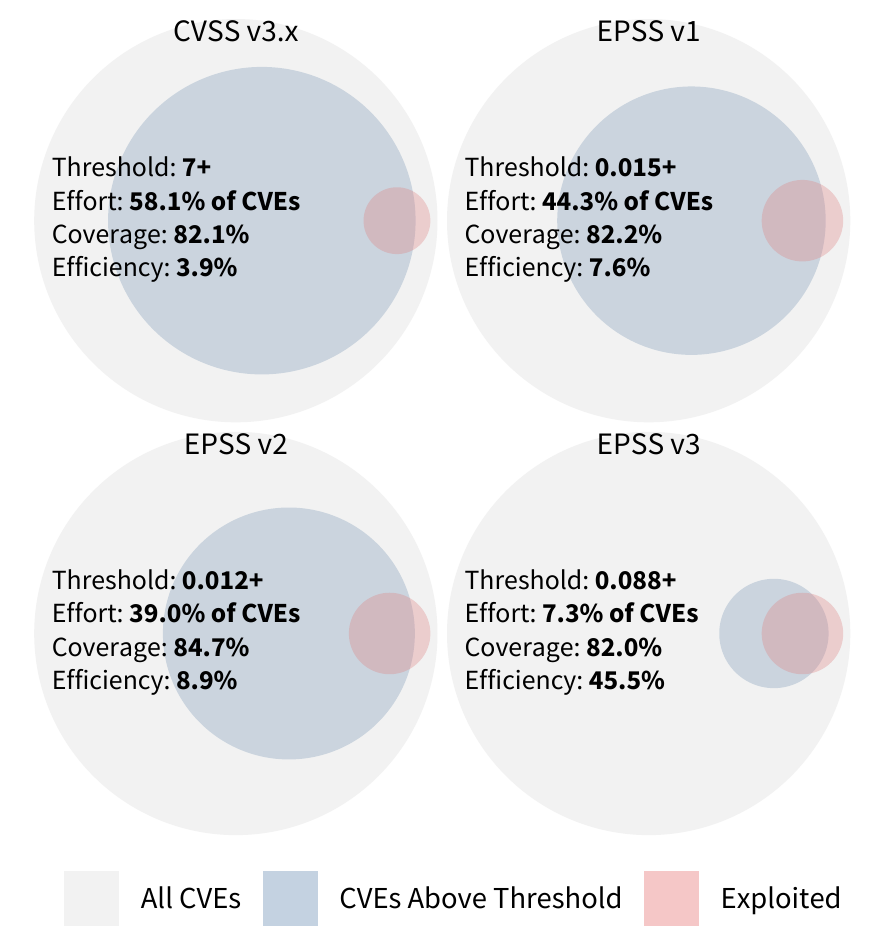}
   \caption{Strategy comparisons holding the coverage constant}
   \label{figure:venn_holding_coverage}
\end{figure}

Figure \ref{figure:venn_holding_coverage} compares the four strategies while maintaining approximately the same level of coverage. That is, the proportion of the red circle (exploitation activity) covered by the blue circle (number of vulnerabilities needing to be remediated). The baseline for coverage is set by a CVSS strategy of remediating vulnerabilities with a base score of 7 and above (CVEs with a "High" or "Critical" CVSS score). Such a strategy yields a respectable coverage at 82.1\% but at the cost of a higher level of effort, needing to remediate 58.1\% or 110,000 of all published CVEs. Practitioners can achieve a similar level of coverage (82\%) using EPSS v3 and prioritizing vulnerabilities scored at 0.088 and above but with a much lower level of effort, needing to only remediate 7.3\% or just under 14,000 vulnerabilities.

\noindent\fbox{%
    \parbox{\linewidth}{%
    Remediating CVEs rated as High or Critical with CVSS v3 gives a respectable level of coverage at 82.1\%, but requires remediating 58.1\% of published CVEs. On the other hand, EPSS v3 can achieve the same level of coverage but reduces the amount of effort from 58.1\%  to 7.3\% of all CVEs, or fewer than 14000 vulnerabilities. 
    }%
}

\section{Discussion and Future Work}
\label{section:discussion}
Currently, the EPSS model ingests data concerning which vulnerabilities were exploited on which days. However, exploitation has many other characteristics, which may be useful to capture and examine. For example, we may be interested in studying the number of exploits per vulnerability (volume), fragmentation of exploitation over time (that is, the pattern of periods of exploitation), or prevalence, which would measure the spread of exploitation, typically by counting the number of devices detecting exploitation. We leave these topics for future work. 

\subsection{Comparison to neural networks}\label{section:neural-net}

In addition to the XGBoost model presented in \autoref{section:evaluation}, we also train a transformer-based classifier on our data set. Transformers~\cite{vaswani2017attention} have achieved state-of-the-art performance in a wide range of sequence modeling tasks, especially for natural language processing. Note that our feature set can be thought of as a sequence of tag/value pairs $(t_i, v_i)$ that have been assigned to a CVE, where $t_i$ contains the integer index assigned to a tag, and $v_i$ represents the associated value (e.g., a count, or simply one for binary features).\footnote{We normalize values associated with each feature/tag to have a maximum of one.} To feed this sequence to a transformer model, we convert each item/tag to an $n$-dimensional embedding using $f_\theta(t_i, v_i) := e_\theta(t_i) + g_\theta(v_i)$, where $e_\theta(\cdot)$ is an embedding lookup table, and $g_\theta(\cdot)$ maps a value to an $n$-dimensional embedding. We use $n=256$, 4 layers, 4 attention heads, and an intermediate layer size of 1024. For $g_\theta(\cdot)$, we use a fully connected neural network with two layers, a hidden layer size of 256, and the tanh activation function.

We train the above classifier for 100,000 iterations with a batch size of 128 and a learning rate of 0.0001, achieving a precision-recall AUC of 0.7374 (as opposed to 0.7795 for the XGBoost model presented in \autoref{section:evaluation}). We believe the slightly lower performance to be due to the aptness of XGBoost for modeling tabular data, and lower susceptibility to overfitting. This further justifies our original model choice for predicting exploitation in-the-wild.

\subsection{Limitations and adversarial consideration}

This research is conducted with a number of limitations. First, insights are limited to data collected from our data partners and the geographic and organizational coverage of their network collection devices. While these data providers collectively manage hundreds of thousands of sensors across the globe, and across organizations of all sizes and industries, they do not observe every attempted exploit event in every network. Nevertheless, it is plausible to think that the data used, and therefore any inferences provided, are representative of all mass exploitation activity.

In regard to the nature of how vulnerabilities are detected, any signature-based detection device is only able to alert on events that it was programmed to observe. Therefore, we are not able to observe vulnerabilities that were exploited but undetected by the sensor because a signature was not written. 

Moreover, the nature of the detection devices generating the events will be biased toward detecting network-based attacks, as opposed to attacks from other attack vectors such as host-based attacks or methods requiring physical proximity.\footnote{For example, it is unlikely to find evidence of exploitation for CVE-2022-37418 in our data set, a vulnerability in the remote keyless entry systems on specific makes and models of automobiles.} Similarly, these detection systems will be typically installed on public-facing perimeter internet devices, and therefore less suited to detecting computer attacks against internet of things (IoT) devices, automotive networks, ICS, SCADA, operational technology (OT), medical devices, etc. 

Given the exploit data from the data partners, we are not able to distinguish between exploit activity generated by researchers or commercial entities, versus actual malicious exploit activity. While it is likely that some proportion of exploitation does originate from non-malicious sources, at this point we have no reliable way of estimating the true proportion. However, based on the collective authors' experience, and discussions with our data providers, we do not believe that this represents a significant percentage of exploitation activity. 

While these points may limit the scope of our inferences, to the extent that our data collection is representative of an ecosystem of public-facing, network-based attacks, we believe that many of the insights presented here are generalizable beyond this dataset.

In addition to these limitations, there are other adversarial considerations that fall outside the scope of this paper. For example, one potential concern is the opportunity for adversarial manipulation either of the EPSS model, or using the EPSS scores. For example, it may be possible for malicious actors to poison or otherwise manipulate the input data to the EPSS model (e.g. Github, Twitter). These issues have been studied extensively in the context of machine learning for exploit prediction~\cite{sabottke2015vulnerability} and other tasks~\cite{suciu2018does,chakraborty2018adversarial}, and their potential impact is well understood. Given that we have no evidence of such attacks in practice, and our reliance on data from many distinct sources which would reduce the leverage of adversaries, we leave an in-depth investigation of the matter for future work. Additionally, it is possible that malicious actors may change their strategies based on EPSS scores. For example, if network defenders increasingly adopt EPSS as the primary method for prioritizing vulnerability remediation, thereby deprioritizing vulnerabilities with lower EPSS scores, it may be conceivable that attackers begin to strategically incorporate these lower scoring vulnerabilities into their tactics and malware. While possible, we are not aware of any actual or suggestive evidence to this effect. 

Finally, while evolving the model from a logistic regression to a more sophisticated machine learning approach greatly improved performance of EPSS, an important consequence is that interpretability of variable contributions is more difficult to quantify as we discuss in the next section.    

\subsection{Variable importance and contribution}

While an XGBoost model is not nearly as intuitive or interpretable as linear regression, we can use SHAP values~\cite{lundberg2017unified} to reduce the opacity of a trained model by quantifying feature contributions, breaking down the score assigned to a CVE as $\phi_0 + \sum_i \phi_i$, where $\phi_i$ is the contribution from feature $i$, and $\phi_0$ is a bias term. We use SHAP values due to their good properties such as local accuracy (attributions sum up to the output of the model), missingness (missing features are given no importance), and consistency (modifying a model so that a feature is given more weight never decreases its attribution). 

\begin{figure}[t]
	\centering
	\includegraphics[width=1\columnwidth]{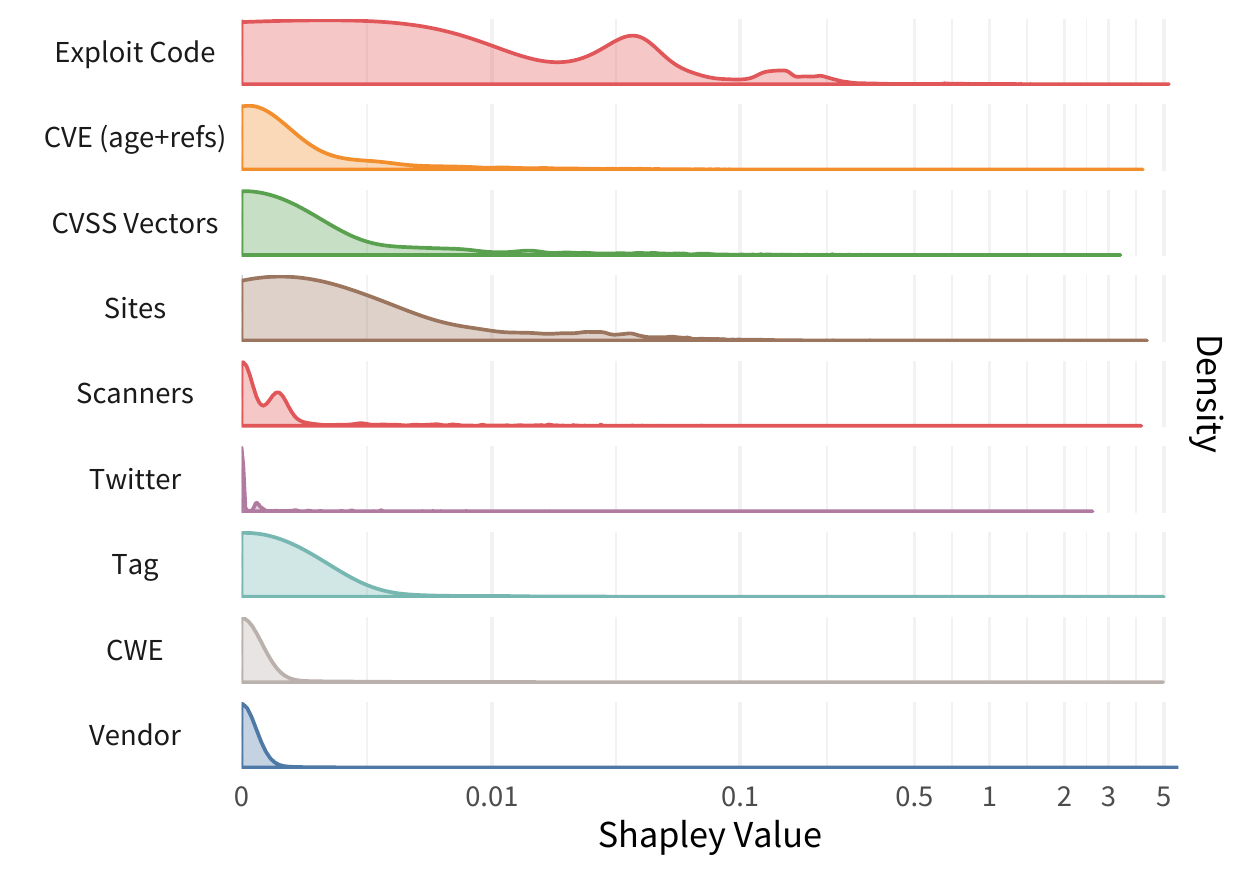}
	\caption{Density plots of the absolute SHAP values for each family of features}
	\label{figure:aggregated-contribution}
\end{figure}

The contributions from different classes of variables in the kernel density plot are shown in Figure \ref{figure:aggregated-contribution}.  First, note that the figure displays the absolute value of the SHAP values, in order to infer the contribution of the variable away from zero. Second, note the horizontal axis is presented on log scale to highlight that the majority of features do not contribute much weight to the final output. In addition, the thin line extending out to the right in Figure \ref{figure:aggregated-contribution} illustrates how there are instances of features within each class that contribute a significant amount. Finally, note that Figure \ref{figure:aggregated-contribution} is sorted in decreasing mean absolute SHAP value for each class of features, highlighting the observation that published exploit code is the strongest contributor to the estimated probability of exploitation activity.

\begin{figure}[t]
	\centering
	\includegraphics[width=1\columnwidth]{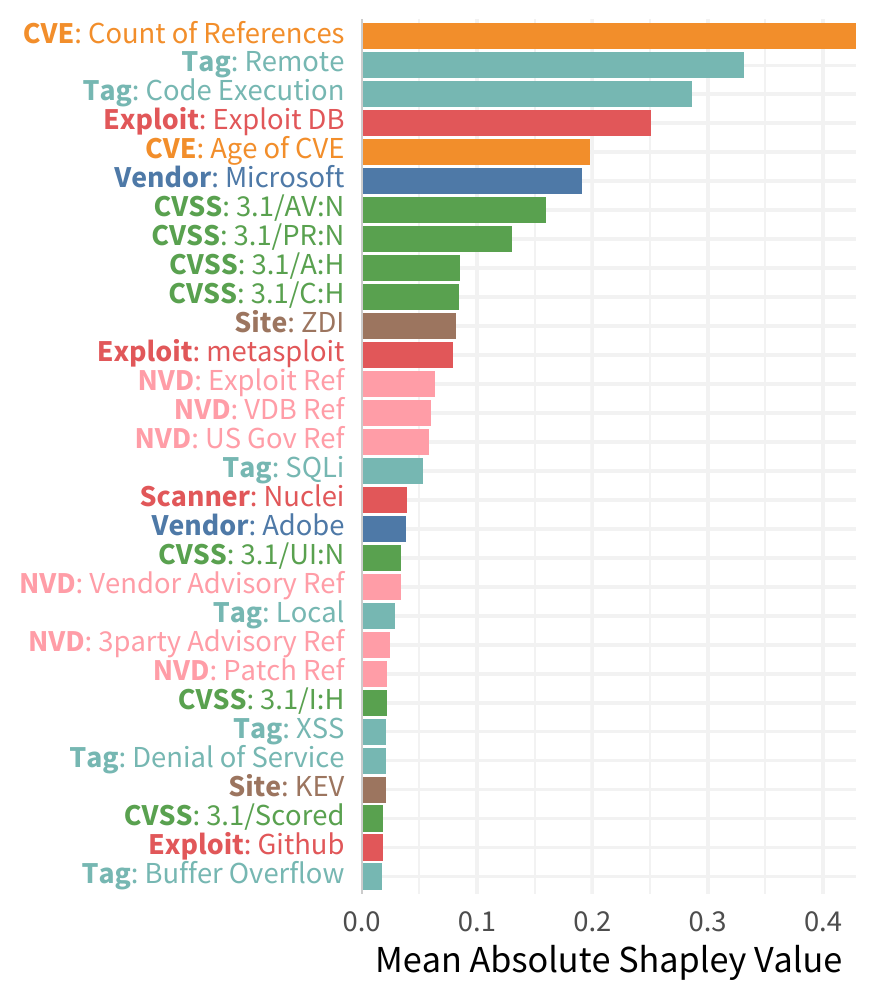}
	\caption{Mean absolute SHAP value for individual features}
	\label{figure:individual-contribution}
\end{figure}

Figure \ref{figure:individual-contribution} identifies the 30 most significant features with their calculated mean absolute SHAP value. Again, note that higher values infer a greater influence (either positive or negative) on the final predicted value. Note that Figure \ref{figure:aggregated-contribution} is showing the mean absolute SHAP value from an entire class of features. So even though Exploit Code as a class of features has a higher mean absolut SHAP value, the largest individual feature is coming from the count of references in the published CVE (which is in the "CVE" class).

\noindent\fbox{%
    \parbox{\linewidth}{%
Note how the most influential feature is the count of the number of references in MITRE's CVE List, followed by ``remote attackers,'' ``code execution,'' and published exploit code in Exploit-DB, respectively. 
    }%
}

\section{Literature Review and Related Scoring Systems}

This research is informed by multiple bodies of literature. First, there are a number of industry efforts that seek to provide some measure of exploitability for individual vulnerabilities, though there is wide variation in their scope and availability. First, the base metric group of CVSS, the leading standard for measuring the severity of a vulnerability, is composed of two parts, measuring impact and exploitability~\cite{cvss3guide}.
The score is built on expert judgements, capturing, for example the observation that a broader ability to exploit a vulnerability (i.e., remotely across the Internet, as opposed to requiring local access to the device); a more complex exploit required, or more user interaction required, all serve to increase the apparent likelihood that a vulnerability could be exploited, all else being equal. CVSS has been repeatedly shown by prior work~\cite{allodi2012preliminary,allodi2014comparing}, as well as our own evidence, to be insufficient for capturing all the factors that drive exploitation in the wild. The U.S. National Vulnerability Database (NVD) includes a CVSS base score with nearly all vulnerabilities it has published. Because of the wide-spread use of CVSS, specifically the base score, as a prioritization strategy we will compare our performance against CVSS as well as our previous models. 

Exploit likelihood is also modeled through various vendor-specific metrics. In 2008, Microsoft introduced the Exploitability Index for vulnerabilities in their products~\cite{MS:ExploitabilityIndex}.
It provides 4 measures for the likelihood that a vulnerability will be exploited: whether an exploitation has already been detected, and whether exploitation is more or less likely, or unlikely. The metric has been investigated before~\cite{reuters2009microsoft,eiram2013exploitability,younis2015comparing} and was shown to have limited performance at predicting exploitation in the wild~\cite{MS:ExploitabilityIndexCritiqueDarkReading,reuters2009microsoft} or the development of functional exploits~\cite{suciu2022expected}.

Redhat provides a 4-level severity rating: low, moderate, important, and critical~\cite{RedHat:SeverityRating}.
In addition to capturing a measure of the impact to a vulnerable system, this index also captures some notion of exploitability. For example, the ``low'' severity rating represents vulnerabilities that are unlikely to be exploited, whereas the ``critical'' severity rating reflects vulnerabilities that could be easily exploited by an unauthenticated remote attacker. Like the Exploitability Index, Redhat's metric is vendor-specific and has limitations reflecting exploitation likelihood~\cite{suciu2022expected}.

A series of commercial solutions also aim to capture the likelihood of exploits. Tenable, a leading vendor of intrusion detection systems, created the Vulnerability Priority Rating (VPR), which, like CVSS, combines information about both impact to a vulnerable system, and the exploitability (threat) of a vulnerability in order to help network defenders better prioritize remediation efforts~\cite{VPRDoc}.
For example, the threat component of VPR ``reflects both recent and potential future threat activity'' by examining whether exploit code is publicly available, whether there are mentions of active exploitation on social media or in the dark web, etc.  Rapid 7's Real Risk Score product uses its own collection of data feeds to produce a score between 1-1000.
This score is a combination of the CVSS base score, ``malware exposure, exploit exposure and ease of use, and vulnerability age'' and seeks to produce a better measure of both exploitability and ``risk''~\cite{Rapid7Risk}. Recorded Future's Vulnerability Intelligence product integrates multiple data sources, including threat information, and localized asset criticality~\cite{recordedfuturerisk}.
The predictions, performance evaluations and implementation details of these solutions are not publicly available. 

These industry efforts are either vendor-specific, score only subsets of vulnerabilities, based on expert opinion and assessments and therefore not entirely data-driven, or proprietary and not publicly available.

Our work is also related to a growing academic research field of predicting and detecting vulnerability exploitation. A large body of work focuses on predicting the emergence of proof-of-concept or functional exploits ~\cite{bozorgi2010beyond,edkrantz2015predicting,bullough2017predicting,reinthal2018data,alperin2019risk,bhatt2021exploitability,suciu2022expected}, not necessarily whether these exploits will be used in the wild, as is done with EPSS. Papers predicting exploitation in the wild have used alternative sources of exploitation, most notably data from Symantec's IDS, to build prediction models~\cite{sabottke2015vulnerability,almukaynizi2017proactive,chen2019using,xiao2018patching,tavabi2018darkembed,fang2020fastembed,hoque2021improved}). Most of these papers build vulnerability feature sets from commonly used data sources such as NVD or OSVDB, although some of them use novel identifiers for exploitation: \cite{sabottke2015vulnerability} infers exploitation using Twitter data, \cite{xiao2018patching} uses patching patterns and blacklist information to predict whether organizations are facing new exploits, while \cite{tavabi2018darkembed} uses natural language processing methods to infer context of darkweb/deepweb discussions. 

\noindent\fbox{%
    \parbox{\linewidth}{%
Compared to other scoring systems and research described above, EPSS is a rigorous and ongoing research effort is; an international, community-driven effort; designed to predict vulnerability exploitation in the wild; available for all known and published vulnerabilities;  updated daily to reflect new vulnerabilities and new exploit-related information; made available freely to the public. 
    }%
}

\section{Conclusion}
In this paper, we presented results from an international, community-driven effort to collect and analyze software vulnerability exploit data, and to build a machine learning model capable of estimating the probability that a vulnerability would be exploited within 30 days following the prediction. In particular, we described the process of collecting each of the additional variables, and described the approaches used to create the machine learning model based on 6.4 million observed exploit attempts. Through the expanded data sources we achieved an unprecedented 82\% improvement in classifier performance over the previous iterations of EPSS.

We illustrated practical use of EPSS by way of comparison with a set of alternative vulnerability remediation strategies. In particular, we showed the sizeable and meaningful improvement in coverage, efficiency and level of effort (as measured by the number of vulnerabilities that would need to be remediated) by using EPSS v3 over any and all current remediation approaches, including CVSS, CISA's KEV list, and Metasploit. 

As the EPSS effort continues to grow, acquire and ingest new data, and improve modeling techniques with each new version, we believe it will continue to improve in performance, and provide new and fundamental insights into vulnerability exploitation for many years to come.

\section*{Acknowledgements} We would like to acknowledge the participants of the EPSS Special Interest Group (SIG), as well as the organizations that have contributed to the EPSS data model to include: Fortinet, Shadow Server Foundation, Greynoise, Alien Vault, Cyentia, and FIRST. 

\bibliographystyle{ACM-Reference-Format}
\bibliography{main}

\end{document}